\documentclass[showpacs,aps,prd,floatfix,amsmath,amssymb,nofootinbib,superscriptaddress]{revtex4-2}
\usepackage{graphicx}
\usepackage{amssymb}
\usepackage{bbm}
\usepackage{hyperref}
\usepackage{color}
\usepackage{slashed}
\usepackage{float}
\usepackage{subfigure}
\usepackage{dcolumn}
\usepackage{hyperref}

\usepackage{multirow}
\begin{document}

\title{ $H\rightarrow Z\gamma$ decay and $CP$ violation}
\author{A. I. Hern\'andez-Ju\'arez}
\email{alan.hernandez@cuautitlan.unam.mx}
\affiliation{Departamento de F\'isica, FES-Cuautitl\'an, Universidad Nacional Aut\'onoma de M\'exico, C.P. 54770, Estado de M\'exico, M\'exico.}
\author{R. Gait\'an}
\affiliation{Departamento de F\'isica, FES-Cuautitl\'an, Universidad Nacional Aut\'onoma de M\'exico, C.P. 54770, Estado de M\'exico, M\'exico.}
\author{ R. Martinez}
\affiliation{Departamento de F\'isica, Universidad Nacional de Colombia, K. 45 No. 26-85, Bogot\'a D.C., Colombia}
\date{\today}


\date{\today}

\begin{abstract}
This study examines the impact of $CP$-violation on the signal strength $\mu^{Z\gamma}$, which was reported as $2.2\pm 0.7$ by the LHC. We obtain constraints on the real and absorptive parts of the $CP$-violating form factor $h_3^{Z\gamma}$ and find that they are less than 1.15 GeV. Additionally, we revisit the leading order Standard Model contributions to the $H\rightarrow Z\gamma$ decay and calculate contributions to $h_3^{Z\gamma}$ from FCNC complex couplings mediated by the $Z$ and $H$ bosons. By using the current bounds on such couplings, we find that the FCNC contribution to $h_3^{Z\gamma}$ with top and charm quarks in the loop is of order $10^{-5}$ GeV. While in a model with new quarks that preserves the SM predictions on Higgs decays,  the $CP$-violating form factor $h_3^{Z\gamma}$ can be of order $10^{-1}$ GeV and could explain the excess on the signal strength $\mu^{Z\gamma}$.

\end{abstract}


\date{\today}

\maketitle
\section{ Introduction}
\label{intro}
 
The LHC measurement of the Higgs boson in 2012 \cite{CMS:2012qbp, ATLAS:2012yve} is considered one of the most significant achievements in particle physics, representing the completion of the Glashow-Weinberg-Salam Standard Model (SM) \cite{GLASHOW1961579, PhysRevLett.19.1264,osti_4767615}. Subsequent measurements have consistently corroborated the Higgs couplings as predicted by the SM  \cite{Aad:2016aa, Aad:2022aa, Tumasyan:2022aa}. Nonetheless, ongoing studies at the LHC persist in thoroughly probing its properties and decay modes. For instance, two years ago, the CMS collaboration released the first measurements of the Higgs boson width and the off-shell contributions to $ZZ$ production \cite{CMS:2022ley}. These findings were also reported by the ATLAS collaboration last year \cite{ATLAS:2023dnm}. Furthermore, both collaborations also reported  the production of a $Z\gamma$ pair through a Higgs boson \cite{CMS:2022ahq,ATLAS:2023yqk}, with the measured signal yield being
\begin{equation}
\label{signalLHC}
\mu^{Z\gamma}=2.2\pm 0.7,
\end{equation}
indicating that the observed rate in the $H\rightarrow Z\gamma$ decay is twice the rate expected in the SM. This discrepancy suggests the potential involvement of new physics phenomena, which have been studied  through  lepton polarizations \cite{Ahmed:2023vyl}, the introduction of new particles \cite{Barducci:2023zml,Lichtenstein:2023vza,Boto:2023bpg,Cheung:2024kml,He:2024sma}, in the context of the minimal left-right symmetric model \cite{Hong:2023mwr}, minimal supersymmetric SM \cite{Israr:2024ubp}, non-supersymmetric models \cite{Benbrik:2022bol}, two loop contributions \cite{Chen:2024vyn,Sang:2024vqk}, Two Higgs Doublet Model \cite{Panghal:2023iqd,Chen:2024vyn} and considering SM interference terms \cite{Buccioni:2023qnt}. 
 \begin{figure}[H]
\begin{center}
\includegraphics[width=8cm]{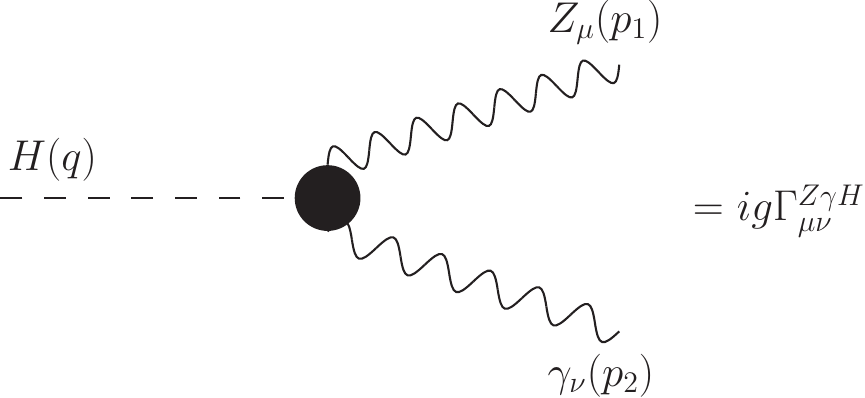}
\caption{$HZ\gamma$ coupling} \label{Vertex}
\end{center}
\end{figure}
 The $H\rightarrow Z\gamma$ decay 
is induced by loops in the SM \cite{Hue:2017cph}. The vertex function for this process can be written as
\begin{equation}
\label{VertexFunction}
\Gamma_{Z\gamma H}^{\mu\nu}=h_1^{Z\gamma} g^{\mu\nu}+\frac{1}{m_Z^2}\Big\{ h^{Z\gamma}_2 p_1^\nu p_2^\mu+h^{Z\gamma}_3\epsilon^{\mu\nu\alpha\beta}p_{1\alpha}p_{2\beta}\Big\},
\end{equation}
where we use the notation and kinematics shown in Fig. \ref{Vertex}.  The form factors $h^{Z\gamma}_1$ and $h^{Z\gamma}_2$ are $CP$-conserving and are induced at the one-loop level in the SM. On the other hand, the  $h_3^{Z\gamma}$ form factor is $CP$-violating.  It is important to note that introducing the $1/m_Z^2$ factor in Eq. \eqref{VertexFunction} implies that all the form factors have the units of mass. 
Due to gauge invariance, the $h^{Z\gamma}_1$ and $h^{Z\gamma}_2$ form factors are not independent:
\begin{equation}\label{h2toh1}
h^{Z\gamma}_2=\frac{2\  m^2_Z}{m_Z^2-m_H^2}h^{Z\gamma}_1.
\end{equation}
Then, by using Eq. \eqref{VertexFunction}, we can express the partial width $\Gamma(H\rightarrow Z\gamma)$ in terms of only two forms factors, which can be written as
\begin{align}
\label{WidthGeneral}
\Gamma(H\rightarrow Z\gamma)&=g^2\frac{m_H^2-m_Z^2}{32\ \pi m_H^3 m_Z^4} \Big( 4 | h^{Z\gamma}_1|^2m_Z^4+ |h^{Z\gamma}_3|^2\big(m_H^2-m_Z^2\big)^2 \Big)\nonumber\\
&= \Gamma^{\text{SM}}(H\rightarrow Z\gamma)+ \delta\Gamma(H\rightarrow Z\gamma),
\end{align}
where the correction arising from the $CP$-violating form factor is given by
\begin{equation}\label{delta}
\delta\Gamma(H\rightarrow Z\gamma)= g^2\frac{\big(m_H^2-m_Z^2\big)^3 }{32\ \pi m_H^3 m_Z^4} |h^{Z\gamma}_3|^2.
\end{equation}

The SM contributions to $h_{1,2}^{Z\gamma}$ were studied in Refs. \cite{Cahn:1978nz,Bergstrom:1985hp,Gunion:1989we,Decker:1991dz}. The QCD corrections have been calculated numerically \cite{Spira:1991tj} and analytically \cite{Bonciani:2015aa, Gehrmann:2015aa}. Additionally, new physics contributions to $h_{1,2}^{Z\gamma}$ have also been revisited in the context of extended scalar sectors \cite{Fontes_2014, Yildirim_2022, Hue_2023}, 331 models \cite{Yue_2013, Hung_2019}, SM effective field theory \cite{Bellazzini_2018, Dedes_2019}, supersymmetric models \cite{WEILER1989337, Cao_2013, Hammad_2015, Liu_2020, Archer_Smith_2021}, new fermions \cite{Bizot_2016} and in left-right symmetric models \cite{Martinez:1989bg, Martinez:1990ye, MARTINEZ1990503}. Furthermore, the $H\rightarrow Z\gamma$ decay has been incorporated into publicly available codes that include higher order corrections \cite{Djouadi_1998, Djouadi_2019hdec,Heinemeyer_2000}. The study of the sensitivity of the $H\rightarrow Z\gamma$ signal strength to new physics has been addressed in Ref. \cite{Cao:2018cms}. Whereas the physics of $h_3^{Z\gamma}$ have been studied in Refs \cite{Senol:2014naa,Hagiwara:2000tk,Rindani:2009pb,Hankele:2006ma,Chen:2017plj,Korchin:2013jja,Feng:2021izk,TaheriMonfared:2016gua,He:2020suf}.

This study explores the possibility that the discrepancy in Eq. \eqref{signalLHC} can be attributed to $CP$-violating effects in the $H\rightarrow Z\gamma$ decay. Assuming this scenario, it is possible to obtain bounds on the $CP$-violating  parameters. Additionally, we examine a model with new quarks and complex FCNC couplings that can induce the $CP$-violating form factor. This model could address the observed difference in the $\mu^{ Z\gamma}$ signal strength while remaining consistent with the experimental results for the $gg\rightarrow H$ and $H\rightarrow \gamma\gamma$ processes.

The structure of this work is organized as follows: we revisit the SM contributions to the $H\rightarrow Z\gamma$ decay in Sec. \ref{SMsec}. In Sec. \ref{Boundh3Sec}, we obtain limits on the $CP$-violating form factor $h_3^{Z\gamma}$, and in Sec. \ref{NPsec}, the $CP$-violating contribution is calculated within models with complex FCNC couplings and new fermions. Finally, our results are summarized in Sec. \ref{Concsec}.

\section{SM contribution}\label{SMsec}

\begin{figure}[H]
\begin{center}
\subfigure[]{\includegraphics[width=5cm]{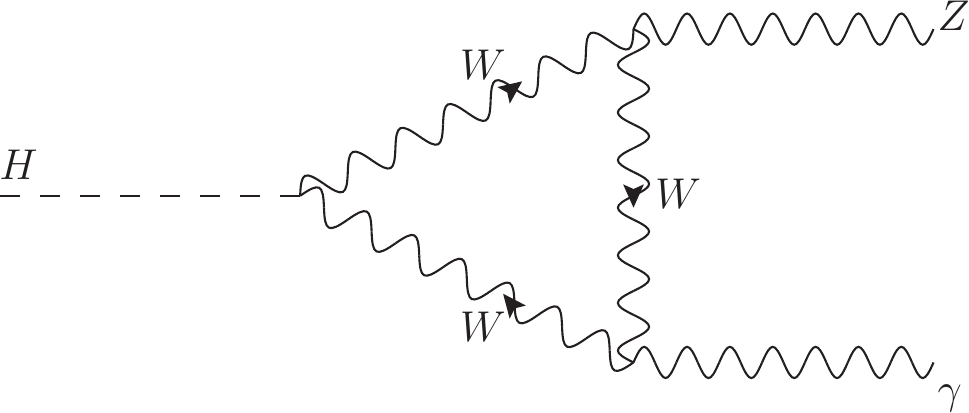}\label{diag1}}
\subfigure[]{\includegraphics[width=5cm]{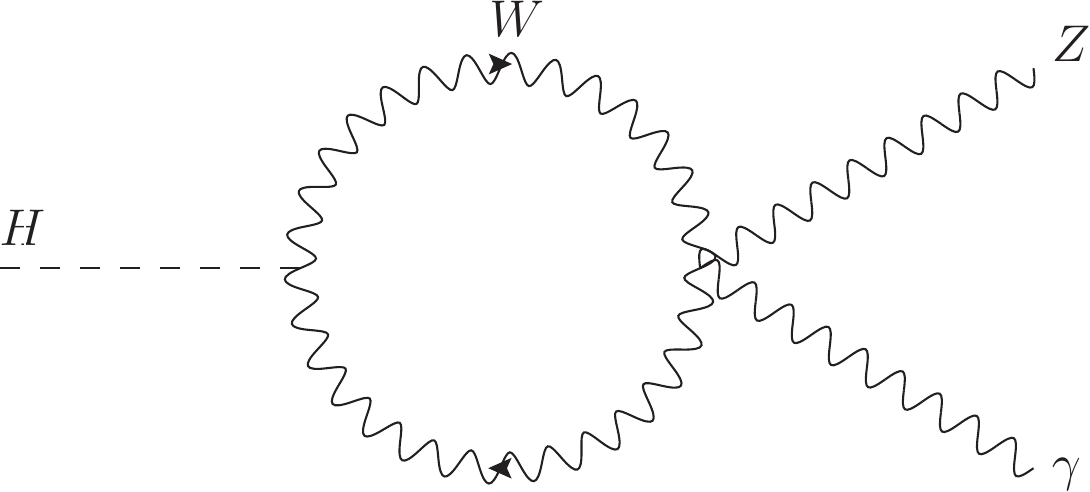}\label{diag2}}
\subfigure[]{\includegraphics[width=5cm]{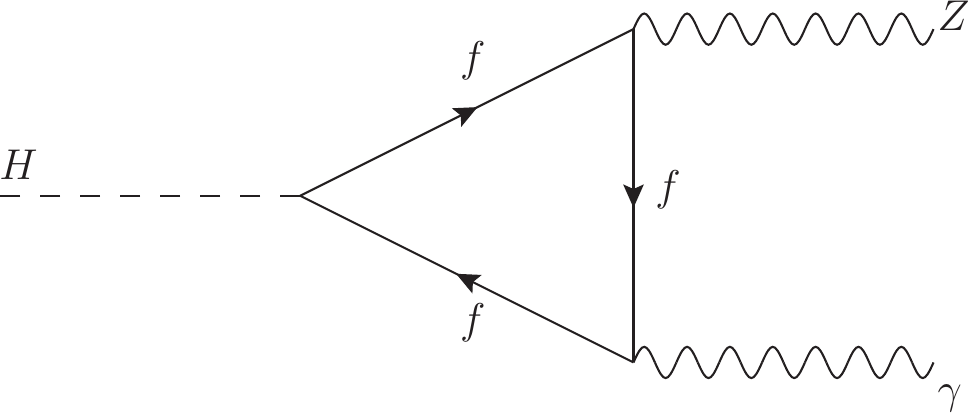}\label{diag3}}
\caption{One-loop SM contributions in the unitary gauge to the $H\rightarrow Z\gamma$ decay. } \label{SMContributions}
\end{center}
\end{figure}

In the unitary gauge, the one-loop SM contributions to the $H\rightarrow Z\gamma$ decay are depicted in Fig. \ref{SMContributions}. The diagrams \ref{diag1}-\ref{diag2} and \ref{diag3} correspond to the $W$ boson ($h_1^{Z\gamma}(W)$) and fermion ($h_1^{Z\gamma}(F)$) loops contributions, respectively. The light fermions coupled to the Higgs boson in diagram \ref{diag3} can be on-shell, leading to an imaginary component in the amplitude \cite{Hernandez-Juarez:2021xhy}. It is worth noting that this imaginary part can be larger than the real one in off-shell couplings \cite{Hernandez-Juarez:2020drn, Hernandez-Juarez:2021mhi, Hernandez-Juarez:2020gxp, He:2020suf}. Furthermore, the $CP$ violation and the imaginary contributions can induce new left-right asymmetries, which could be observed at the LHC \cite{Hernandez-Juarez:2023dor, Godbole_2007}. Although the absorptive part has been discussed in some studies \cite{Gehrmann:2015aa}, its numerical value has been overlooked in recent SM calculations \cite{Hue:2017cph, Degrande:2017naf}. Therefore, we reassess the SM contribution to the $HZ\gamma$ coupling to determine the magnitude of its imaginary part.

With the help of FeynArts \cite{Hahn:2000kx} and FeynCalc \cite{Mertig:1990an, Shtabovenko:2016sxi, Shtabovenko:2020gxv}, we calculated the Feynman diagrams in Fig. \ref{SMContributions}. The total contribution can be written as 
\begin{equation}
h_1^{Z\gamma}=h_1^{Z\gamma}(F)+h_1^{Z\gamma}(W).
\end{equation}
The SM contributions have been reported in Refs. \cite{Cahn:1978nz, Bergstrom:1985hp, KNIEHL1994211, Djouadi:1996yq, Hue:2017cph}. The fermion contribution, in terms of the Passarino-Veltman scalar functions, is given as
\begin{align}
\label{h1F}
h_1^{Z\gamma}(F)=&\frac{N_c \mathcal{Q}_f\ g\ g_V\ e\ m_f^2 }{8 \pi ^2  c_W m_W
   \left(m_H^2-m_Z^2\right)} \Bigg\{2
   m_Z^2\Big[
   \text{B}_0\left(m_H^2,m_f^2,m_f^2
   \right)-
   \text{B}_0\left(m_Z^2,m_f^2,m_f^2
   \right)\Big]\nonumber\\
   &+\left(m^2_H-m^2_Z\right)
   \left(\left(-m_H^2+4
   m_f^2+m_Z^2\right)
   \text{C}_0\left(0,m_H^2,m_Z^2,m_f
   ^2,m_f^2,m_f^2\right)+2\right)\Bigg\},
\end{align}
where $m_f$ and $\mathcal{Q}_f$ denote the mass and charge of the fermion $f$ in the loop, respectively. Additionally, $g_V$ corresponds to the vector coupling of the $Z$ boson with the fermions, and $N_c$ stands for the color number. The contribution from the $W^\pm$ boson loop contribution is
\begin{align}
\label{h1W}
h_1^{Z\gamma}(W)=&-\frac{ g c_W e}{32
   \pi ^2 m_W^3}
   \Bigg\{\frac{2 m_W^2
   \left(m_H^2+6 m_W^2\right)-m_Z^2
   \left(m_H^2+2 m_W^2\right)}{\left(m^2_H-m^2_Z\right)}
   \Big[m_Z^2
   \Big(\text{B}_0\left(m_H^2,m_W^2
   ,m_W^2\right)-\text{B}_0\left(m_Z
   ^2,m_W^2,m_W^2\right)\Big)\nonumber\\
   &+\left(m^2_H-m^2_Z\right)\Big]+2 m_W^2
   \Big[m_H^2 \left(m_Z^2-6
   m_W^2\right)+6 m_W^2 m_Z^2+12
   m_W^4-2 m_Z^4\Big]
   \text{C}_0\left(0,m_H^2,m_Z^2,m_W
   ^2,m_W^2,m_W^2\right)\Bigg\}.
\end{align}
Our results agree with those reported in Ref. \cite{Hue:2017cph}. We utilized PackageX to simplify Eq. \eqref{h1W} and express it in the analytical form of Eq. 20 as reported in Ref. \cite{Phan:2021pcc}, which was derived using a general $R_\xi$ gauge. We observe that both Eq. \eqref{h1F} and \eqref{h1W} are free of divergences. Furthermore, as a cross-check, we computed the $h^{Z\gamma}_2$ form factor and confirmed the validity of Eq. \eqref{h2toh1}.  For the evaluation of the Passarino-Veltman scalar functions, we used LoopTools \cite{Hahn:1998yk}.  In perturbation theory, we find that contributions from light fermions are expected to be negligible, as $h_1^{Z\gamma}(F)\sim m_f^2$. However, these corrections must be carefully addressed due to their nonperturbative nature. The one-loop contributions from light quarks can be analyzed through the chiral limit \cite{PhysRevLett.28.1482,LEUTWYLER1974413,Maris_1997}, which is defined within a nonperturbative framework. In QCD, chiral symmetry is dynamically broken \cite{PhysRevC.68.015203,Bhagwat_2006,Bowman_2005}, providing a mechanism for generating quark masses. Within this context, one can define and calculate one-loop corrections of the light quarks. For instance, their chromomagnetic and chromoelectric dipole moments \cite{Chang_2011}, which can be computed in various kinematical regimes and necessitate a nonperturbative approach \cite{PhysRevD.95.034041}. On the other hand, Higgs analyses at the LHC can be sensitive to corrections from bottom and charm quarks \cite{Marquard:2015qpa}. To study their perturbative contribution at leading order (LO), it is necessary to consider the running quark mass $m_q(\mu)$, where $\mu$ represents the renormalization scale \cite{DJOUADI1993255,SPIRA199517}. In this work,  we use the quark masses $m_t=172.57$ GeV, $m_c=1.51$ GeV, and $m_b=4.18$ GeV, as recommended by the LHCHWG \cite{LHCHiggsCrossSectionWorkingGroup:2016ypw,ParticleDataGroup:2022pth}. The numerical contributions of the leptons, top, bottom, and charm quarks are shown in Table \ref{H3contributions}. Public codes, such as \texttt{HDECAY}, have implemented the masses $\overline{m}_q(\mu)$ in the $\overline{\text{MS}}$ scheme to assess the contributions of the bottom and charm quarks \cite{Djouadi_2019hdec}, in line with the LHCHWG recommendations \cite{LHCHiggsCrossSectionWorkingGroup:2016ypw}. The next to leading order (NLO) QCD contributions to the top and bottom diagrams have been calculated in Ref. \cite{Gehrmann:2015aa}, with the perturbative QCD requiring the running quark masses at the $\mu=m_H$ scale.

The leading order SM contributions are:
 \begin{equation}
\label{h1sm}
h_1^{Z\gamma}(F)=\Big(1.86 \times 10^{-2}+i2.6\times 10^{-4}\Big)\ \text{GeV}\text{,}\quad h_1^{Z\gamma}(W)=-3.4\times10^{-1}\ \text{GeV.}
\end{equation}
It is noted that the main contribution to the real part arises from the $W$ boson diagrams, and it is of order $10^{-1}$ GeV. The imaginary part is three orders of magnitude smaller, with the larger contribution coming from the bottom quark. Furthermore, there is destructive interference between the $W$ boson and top contributions. 

The total SM contribution at leading order is given by
\begin{equation}\label{SMZgammaTotal}
h_1^{Z\gamma}=\Big(-3.22 \times 10^{-1}+i2.6\times 10^{-4}\Big)\ \text{GeV}.
\end{equation}
Although the absorptive part is small, it can still have significant consequences on polarized observables \cite{Hernandez-Juarez:2024zpk}, which can be investigated at colliders following the approach used for the $HZZ$ coupling \cite{Ballestrero_2019, Maina_2021, Maina_20212}. The study of these effects is possible as the polarizations of the $Z$ boson and photon are subjects of study at the LHC and other colliders \cite{Aad:2016ab, Aad:2024aa,2015154, PhysRevLett.129.091801, PhysRevLett.129.091801, Aad:2023aa, LHCb:2019vks, LHCb:2021byf,Belle:2006pxp, BaBar:2008okc}. 

At leading order, the partial width is given by
\begin{equation}
\label{WidthSM}
\Gamma
_{\text{LO}}^{\text{SM}}(H\rightarrow Z\gamma)=6.61 \times 10^{-3} \text{MeV}.
\end{equation}

\begin{table}[H] \begin{center}
 \begin{tabular}{cc} Contribution& $h_1^{Z\gamma}(F)$   \\\hline\hline Top quark &  1.9$\times 10^{-2}$ GeV   \\\hline Charm quark & $(-8.07\times 10^{-5}+i\ 3.32\times10^{-5})$ GeV    \\\hline Bottom quark & $(-3.91\times 10^{-4}+i\ 2.20\times10^{-4})$ GeV    \\\hline Leptons &  $(-1.46\times 10^{-5}+i\ 6.3\times10^{-6})$  GeV  \end{tabular}\caption{Different fermion contributions to the form factor $h_1^{Z\gamma}$.}\label{H3contributions}\end{center}
 \end{table}

\section{Bounds on the $CP$-violating form factor $h^{Z\gamma}_3$}\label{Boundh3Sec}

In this section, we derive constraints on the  $CP$-violating form factor $h_3^{Z\gamma}$, which can be considered to be complex:
\begin{equation}
h^{Z\gamma}_3={\rm Re}\big[h^{Z\gamma}_3\big]+i {\rm Im}\big[h^{Z\gamma}_3\big].
\end{equation}
We assume that the new physics contributions to $h_3^{Z\gamma}$ may explain the observed $\mu^{Z\gamma}$ discrepancy reported by CMS and ATLAS. The signal strength $\mu^{Z\gamma}$ is defined as the ratio between the measured Higgs boson rate and its SM prediction \cite{ATLAS:2016neq}:
\begin{equation}
\label{signal3}
\mu_i^f=\frac{\sigma_i \mathcal{B}^{Z\gamma}}{(\sigma_i)_\text{SM}(\mathcal{B}^{Z\gamma})_\text{SM}},
\end{equation}
where $\sigma_i$ represents the production cross section of the $i\rightarrow H$ process and $\mathcal{B}^{Z\gamma}$ corresponds to the branching fraction for the $H\rightarrow Z\gamma$ decay. 
Considering that the new physics only arises from the $HZ\gamma$ coupling, we can express the signal strength $\mu^{Z\gamma}$ as 
 \begin{align}
\label{Ratio}
\mu^{Z\gamma}&\simeq\frac{\mathcal{B}^{\text{SM}}(H\rightarrow Z\gamma)+\delta\Gamma(H\rightarrow Z\gamma)/\Gamma_H}{\mathcal{B}^{\text{SM}}(H\rightarrow Z\gamma)},
\end{align}
where $\mathcal{B}^{\text{SM}}(H\rightarrow Z\gamma)$ denotes the branching ratio of the $H\rightarrow Z\gamma$ decay in the SM, $\delta\Gamma(H\rightarrow Z\gamma)$ is the correction introduced by the $CP$-violating form factor $h^{Z\gamma}_3$ in Eq. \eqref{delta}, and $\Gamma_H$ represents the total Higgs width.  Since the $h^{Z\gamma}_3$ contribution to $\Gamma_H$ is not being considered, Eq. \eqref{Ratio} represents only an approximation of the beyond SM effects on the signal strength $\mu^{Z\gamma}$. This approach has also been adopted in studies of SMEFT contributions to the decays $H\rightarrow Z\gamma$ and $H\rightarrow \gamma\gamma$ \cite{Mantzaropoulos:2024vpe}. 

\begin{figure}[H]
\begin{center}
\subfigure{}\includegraphics[width=8cm]{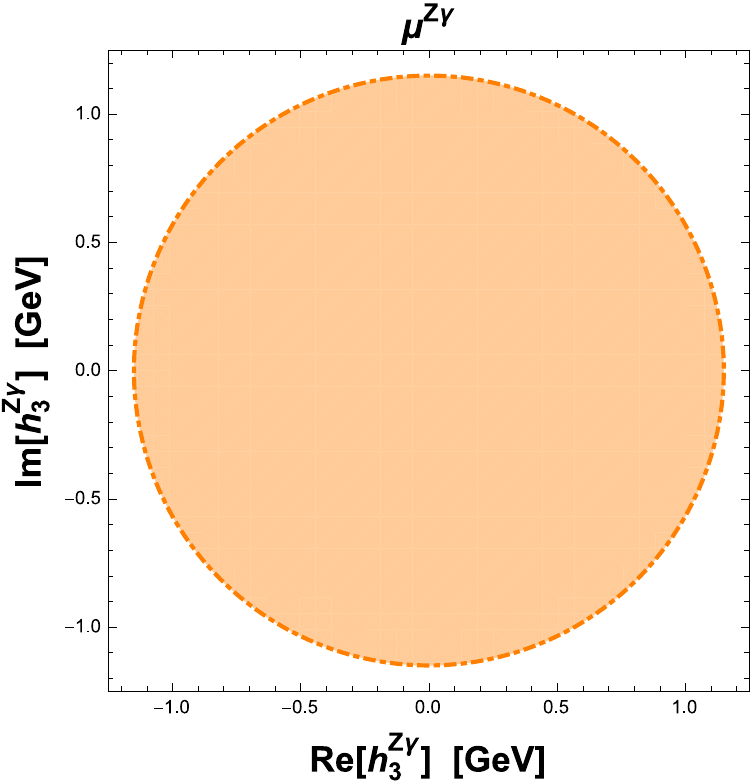}
\caption{Bounds on the real and absorptive parts of the $CP$-violating form factor $h^{Z\gamma}_3$ using the recent value for $\mu^{Z\gamma}$ at the 95\% CL.}\label{Bounds}
\end{center}
\end{figure}

Using the numerical values: $\mathcal{B}^{\text{SM}}(H\rightarrow Z\gamma)= 1.57\times10^{-3}$ \cite{LHCHiggsCrossSectionWorkingGroup:2016ypw} and $\Gamma_H=4.1$ MeV \cite{ParticleDataGroup:2022pth,LHCHiggsCrossSectionWorkingGroup:2016ypw}, we can write the signal strength of the $H\rightarrow Z\gamma$ decay as follows
 \begin{align}
\label{Ratio2}
\mu^{Z\gamma}&=1+(1.94 \text{ GeV}^{-2}) \Big({\rm Re}\big[h^{Z\gamma}_3\big]^2+{\rm Im}\big[h^{Z\gamma}_3\big]^2\Big).
\end{align}
The QCD NLO contributions are not considered, as they are known to be small \cite{Spira:1991tj}. The allowed values in the ${\rm Re}\big[h^{Z\gamma}_3\big]$-${\rm Im}\big[h^{Z\gamma}_3\big]$ plane aligned with the signal strength $\mu^{Z\gamma}$ reported by the LHC can be obtained from Eq. \eqref{Ratio2}. This region is shown in Fig. \ref{Bounds} at the 95\% confidence level (CL), from which the following constraints are obtained:
\begin{equation}\label{limh3}
 \big|{\rm Re}\big[h^{Z\gamma}_3\big] \big|\text{,}\big|{\rm Im}\big[h^{Z\gamma}_3\big] \big|\lesssim 1.15\  \text{GeV}  \quad \text{at 95 \% CL.}
\end{equation}
It is observed that large values of $h_3^{Z\gamma}$ are necessary to explain the anomaly in the $\mu^{Z\gamma}$ signal strength. In previous studies, constraints of the order $10^{-1}-10^{-2}$ GeV have been established in $\gamma q$ induced process \cite{Senol:2014naa} as well as in $e^+e^-$ collisions \cite{Hagiwara:2000tk, Rindani:2009pb}. Angular asymmetries at colliders are a sensitive probe for $h_3^{Z\gamma}$, as they involve the interference between $CP$-even and $CP$-odd couplings \cite{Hankele:2006ma, Chen:2017plj}. The imaginary part of $h_3^{Z\gamma}$ can also introduce new left-right asymmetries, as observed in the $H^\ast ZZ$ vertex \cite{Hernandez-Juarez:2023dor}. Additionally, the effects of $h_3^{Z\gamma}$ have been studied in helicity amplitudes \cite{Korchin:2013jja,Feng:2021izk} and $\gamma$-p collisions \cite{TaheriMonfared:2016gua}.   

\section{New physics contributions}\label{NPsec}

In this section, we calculate the contributions to $h_3^{Z\gamma}$ from a model with complex FCNC couplings. An estimation of its numeric value in a model with SM particles and a new generation of quarks is given.

\subsection{Contributions to $CP$ violation}

 \begin{figure}[H]
\begin{center}
\includegraphics[width=8cm]{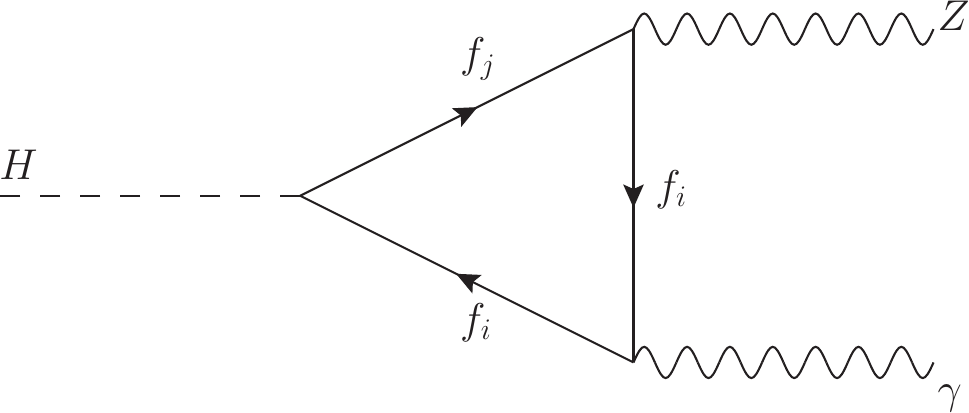}
\caption{One-loop contribution to $h^{Z\gamma}_3$.}\label{FermionesHtc}
\end{center}
\end{figure}

The FCNC couplings of the $Z$ and the Higgs bosons with fermions can be induced from the following Lagrangian 
\begin{align}
\label{lagFCNC}
    \mathcal{L}=\frac{g}{c_W}\overline{f}_i\gamma_\mu\Big(g_V^{ij}-g_A^{ij} \gamma^5\Big)f_j Z^\mu +\frac{g}{2m_W}  \overline{f}_i\Big(g_S^{ij}+g_P^{ij} \gamma^5\Big)f_j H,
\end{align}
where the $g^{ij}_{V\text{,} A}$ and $g^{ij}_{S\text{,} P}$ couplings are considered as complex with the latter having mass dimensions. From the interactions in Lagrangian \eqref{lagFCNC}, the $CP$-violating form factor $h^{Z\gamma}_3$ can be generated through the Feynman diagram shown in Fig. \ref{FermionesHtc}. In terms of the Passarino-Veltman scalar functions, the FCNC contribution to $h_3^{Z\gamma}$ is given as
\begin{align}
\label{h3NP2}
h_3^{Z\gamma}= &\frac{ g\ \mathcal{Q}\ e\ m_Z^2\ N_c}{4 \pi ^2 c_W m_W}\Bigg\{m_j
   \text{C}_0\left(0,m_H^2,m_Z^2,m_j
   ^2,m_j^2,m_i^2\right)\Big[ -{\rm Im }\Big\lbrace g_A^{ij}\left(g_{S}^{ij}\right)^\ast \Big\rbrace+ {\rm Im }\Big\lbrace g_V^{ij}\left(g_{P}^{ij}\right)^\ast \Big\rbrace\Big] 
 \nonumber\\
   &   +m_i
   \text{C}_0\left(0,m_H^2,m_Z^2,m_i
   ^2,m_i^2,m_j^2\right)\Big[ {\rm Im }\Big\lbrace g_A^{ij}\left(g_{S}^{ij}\right)^\ast \Big\rbrace+ {\rm Im }\Big\lbrace g_V^{ij}\left(g_{P}^{ij}\right)^\ast \Big\rbrace\Big]\Bigg\},
\end{align}
where we observe that $h_3^{Z\gamma}$ is free of divergences and depends on the imaginary parts of $g_A^{ij}\left(g_{S}^{ij}\right)^\ast $ and $g_V^{ij}\left(g_{P}^{ij}\right)^\ast $. Furthermore, note that to introduce $CP$ violation in the $HZ\gamma$ vertex, at least one of the four couplings in Lagrangian \eqref{lagFCNC} must be complex. 
 
 When only SM fermions are considered, it is expected that the dominant contributions to $h_3^{Z\gamma}$ will stem from the FCNC couplings between the top and charm quarks. One approach to numerically assess the FCNC contribution to  $h_3^{Z\gamma}$ involves using constraints on the vector and axial couplings of the $Z\overline{t}c$ vertex derived from current LHC data, which are given as \cite{Hernandez-Juarez:2022kjx}
\begin{equation}
\label{BoundsZtc}
|g_{V\text{,}A}^{tc}|<0.0095.
\end{equation}
For the $g_S^{tc}$ and $g_P^{tc}$ couplings, we have derived constraints in Appendix \ref{ABoundHtc}:
 \begin{equation}
\label{BoundsHtc}
|g_{S,P}^{tc}|\lesssim 0.25 \ \text{GeV}.
\end{equation}
For various combinations of the real and imaginary parts of the FCNC couplings within the bounds in Eq. \eqref{BoundsZtc} and \eqref{BoundsHtc}, we find that the absolute value of the real part of $h^{Z\gamma}_3$ can be up the order of $10^{-5}$ GeV. While
the imaginary part of $h^{Z\gamma}_3$ is not induced because the particles coupled to the Higgs boson are not kinematically allowed to be simultaneously on-shell. Our result is significantly lower than the upper bound stated in Eq. \eqref{limh3}, and therefore, can not fully account for the deviation of the signal strength $\mu^{Z\gamma}$ reported by the LHC. However, alternative models beyond the Standard Model could provide additional contributions.  The contributions from FCNC couplings of the down quarks are expected to be small. Thus, they are not considered in this work.

 \subsection{New fermions contributions}
 
Recent works have proposed models with new leptons that align with the observed deviation on the $H\rightarrow Z\gamma$ signal strength \cite{Barducci:2023zml, He:2024sma}.  Under the context of new fermions, we will now consider a hypothetical model that incorporates new up ($T$) and down ($D$) quarks. In this model, the new quarks interact through FCNC couplings of the $Z$ and $H$ bosons with the top and bottom quarks, as outlined in the Lagrangian \eqref{lagFCNC}. 
 It is convenient to express the scalar and pseudo-scalar Higgs couplings in terms of the $m_T$ and $m_D$ masses as follows
 \begin{align}
g^{\overline{Q}q}_{r}= K^{\overline{Q}q}_r m_{Q}\text{,}\quad q=t,b\text{,} \quad Q=T, D\text{, and}\quad r=S,P,
 \end{align}
 while for the vector and axial couplings of the $Z$ boson, we use
 \begin{align}
g^{\overline{Q}q}_{r}= K^{\overline{Q}q}_r\text{,}\quad (r=V,A),
 \end{align}
where, $K^{\overline{Q}q}_r$ ($r=V$, $A$, $S$, $P$) are dimensionless complex constants. Following this notation, the Lagrangian that gives rise to the FCNC couplings of the new quarks with the $Z$ and $H$ bosons is 
 \begin{align} 
\label{lagFCNCQ}
    \mathcal{L}=&\frac{g}{c_W}\overline{Q}\gamma_\mu\Big(K^{\overline{Q}q}_V -K^{\overline{Q}q}_A  \gamma^5\Big)q Z^\mu+\frac{g\ m_{Q}}{2m_W}  \overline{Q}\Big(K^{\overline{Q}q}_S +K^{\overline{Q}q}_P  \gamma^5\Big)q H\nonumber\\
    &-\kappa\frac{g \ m_Q }{2m_W}\overline{Q}Q H.
\end{align}
  In the second line of Eq. \eqref{lagFCNCQ}, we have included the diagonal coupling of the Higgs boson with the new quarks. The effects of a new family of fermions in the Higgs boson physics have been studied in Refs. \cite{Inami:1983aa, Pilaftsis1992, Arhrib:2000ct, Arik:2001iw, Arik:2002ci, Belotsky:2002ym, Arik:2005ed, Kribs:2007aa, Denner:2011vt, Carpenter:2011wb, Englert:2011us, Djouadi:2012ae, Eberhardt:2012gv,Lenz:2013iha}. It is well known that   the new quarks corrections lead to deviations of the $H$ decays into $\gamma\gamma$ and $gg$ from the predictions of the SM \cite{Djouadi:2012ae}. Therefore, to minimize the contributions of $T$ and $D$ quarks in such processes, we have introduced the $\kappa$ constant in the $H\overline{Q}Q$ coupling. By using the notation in Lagrangian \eqref{lagFCNCQ}, the $h_3^{Z\gamma}$ form factor in Eq. \eqref{h3NP2} can be expressed as 
 \begin{align}
\label{h3NP22}
h_3^{Z\gamma}= &\frac{ g\ \mathcal{Q}\ e\ m_Z^2\ N_c}{4 \pi ^2 c_W m_W}\Bigg\{m_Qm_q
   \text{C}_0\left(0,m_H^2,m_Z^2,m_q
   ^2,m_q^2,m_Q^2\right)\Big[ -{\rm Im }\Big\lbrace K_A^{\overline{Q}q}\left(K_{S}^{\overline{Q}q}\right)^\ast \Big\rbrace+ {\rm Im }\Big\lbrace K_V^{\overline{Q}q}\left(K_{P}^{\overline{Q}q}\right)^\ast \Big\rbrace\Big] 
 \nonumber\\
   &   +m_Q^2
   \text{C}_0\left(0,m_H^2,m_Z^2,m_Q
   ^2,m_Q^2,m_q^2\right)\Big[ {\rm Im }\Big\lbrace K_A^{\overline{Q}q}\left(K_{S}^{\overline{Q}q}\right)^\ast \Big\rbrace+ {\rm Im }\Big\lbrace K_V^{\overline{Q}q}\left(K_{P}^{\overline{Q}q}\right)^\ast \Big\rbrace\Big]\Bigg\}.
\end{align} 
 We observe that the parameter space required to evaluate the contribution of the new quarks to $h_3^{Z\gamma}$ consists of six elements: the masses $m_T$ and $m_D$, and the imaginary parts of the products $K_A^{\overline{T}t}\left(K_{S}^{\overline{T}t}\right)^\ast$, $K_V^{\overline{T}t}\left(K_{P}^{\overline{T}t}\right)^\ast$, $K_A^{\overline{D}b}\left(K_{S}^{\overline{D}b}\right)^\ast$ and $K_V^{\overline{D}b}\left(K_{P}^{\overline{D}b}\right)^\ast$. In the rest of this section, we discuss some constraints on the parameter space, determine their allowable values, and use these results to estimate the new contribution to $h_3^{Z\gamma}$.

 \subsubsection{Constraints from SM results}

It is important to consider the impact of the new quarks on the reported signal strength $\mu^{\gamma\gamma}$, which is given by \cite{ParticleDataGroup:2022pth}:
 \begin{align}
\label{HgammaSignal}
   \mu^{\gamma\gamma}=1.10\pm.07.
\end{align}
The decay $H\rightarrow\gamma\gamma$ is influenced by adding the $T$ and $D$ quarks through the fermion loop diagram, as depicted in Fig. \ref{SMContributions}. The vertex function $\Gamma^{H\gamma\gamma}$ can be expressed in a similar form to Eq. \eqref{VertexFunction} (see Appendix  \ref{appendix1}), and the total contribution to the $CP$-conserving form factor $h_1^{\gamma\gamma}$ arises from analogous diagrams in Fig. \ref{SMContributions}. It can be written as
\begin{align}
h_1^{\gamma\gamma}=h_1^{\gamma\gamma}(W)+\kappa\sum_q  h_1^{\gamma\gamma}(F), \quad (q=\ell, t, b, T, D).
\end{align}
   For SM fermions, $\kappa=1$, while for the $T$ and $D$ quarks, it must be less than 1 to reduce their impact on the $H\rightarrow\gamma\gamma$ decay. The expressions for the $h_1^{\gamma\gamma}(W, F)$ form factors can be found in Appendix \ref{appendix1}. In the SM, the contributions from the $W$ and top quark interfere destructively, with the $W$ contribution being dominant. When considering the new quarks, both contributions become comparable but with opposite signs. Therefore, the branching ratio of the $H\rightarrow\gamma\gamma$ decay decreases significantly \cite{Djouadi:2012ae}.
  
To study effects of the $T$ and $D$ quarks on $\mu^{\gamma\gamma}$, we can write the $h_1^{\gamma\gamma}$ form factor as:
\begin{equation}
h_1^{\gamma\gamma}=h_1^{\text{SM}}+h_1^{\text{NP}},
\end{equation}
with $h_1^{\text{SM}}$  and $h_1^{\text{NP}}$ the leading order EW  and new quarks contributions to $h_1^{\gamma\gamma}$, respectively.   The square of the form factor can be written as: 
 \begin{equation}
{\big|h_1^{\gamma\gamma}\big|}^2=\big|h_1^{\text{SM}}\big|^2+\delta \big|{h_1^{\gamma\gamma}}\big|^2,
\end{equation}
where $\delta \big|h_1^{\gamma\gamma}\big|^2= 2{\rm Re}\big( h_1^{\text{SM}} {h_1^{\text{NP}}}^\dagger\big)+\big|h_1^{\text{NP}}\big|^2$. It is noted that $\delta \big|h_1^{\gamma\gamma}\big|^2$ has units of mass squared. 
By using the approach outlined in Sec. \ref{Boundh3Sec} and taking $\mathcal{B}^{\text{SM}}(H\rightarrow \gamma\gamma)=2.27\times 10^{-3}$ \cite{ParticleDataGroup:2022pth}, we can express the signal strength $\mu^{\gamma\gamma}$ as follows
 \begin{align}
\label{HgammaSignal2}
\mu^{\gamma\gamma}&=\frac{\mathcal{B}^{\text{SM}}(H\rightarrow \gamma\gamma)+\delta\Gamma(H\rightarrow \gamma\gamma)/\Gamma_H}{\mathcal{B}^{\text{SM}}(H\rightarrow \gamma\gamma)}\nonumber\\&=1+(7.2\text{ GeV}^{-2} )\delta \big|h_1^{\gamma\gamma}\big|^2,
\end{align}
where $\delta\Gamma(H\rightarrow \gamma\gamma)$ represents the correction arising from $h_1^{\text{NP}}$. It is worth noting that the NLO contributions from ultra-heavy fermions are comparable in magnitude to the LO corrections \cite{Denner:2011vt}. However, we expect these NLO effects to be suppressed for $\kappa<10^{-1}$. The NLO QCD and electro-weak contributions have been calculated using the complex-mass scheme \cite{DENNER2005247,DENNER200622,DENNER199933}, wherein complex $W$ masses are introduced in the two-loop amplitudes in a gauge invariant approach \cite{PASSARINO2007298}. It turns out that both NLO corrections largely cancel each other out \cite{MAIERHOFER2013131}.  Thus, only the LO corrections are considered in our analysis.

The Higgs production through gluon fusion is enhanced by a factor of nine due to contributions of new quarks \cite{Gunion:1994zm, Gunion:1995tp}. Nevertheless, it is anticipated that for small values of $\kappa$, the coupling modifier $\kappa_g$ will not deviate significantly from the recent result obtained by the ATLAS collaboration \cite{ATLAS:2022tnm}:
\begin{equation}
\label{kggh}
\kappa_g=1.01^{+0.11}_{-0.09}.
\end{equation}
When incorporating contributions from the new quarks, the coupling modifier $\kappa_g$ can be written as \cite{Das:2017mnu, Hernandez-Juarez:2018uow}
\begin{align}
\label{kggh2}
\kappa_g^2&=\frac{\Bigg|A_t(\tau_t)+\kappa\sum\limits_f A_f(\tau_f)\Bigg|^2}{\Big| A_t(\tau_t)\Big|^2},\quad f= T,\ D,
\end{align}
where $A_f(\tau_f)$ is defined in Appendix \ref{appendix2}.  In the SM, the bottom quark provides a 10\% of the contribution to the $h\rightarrow gg$ decay \cite{Spira:2019iec}. Nevertheless, considering only the top quark contribution is a reasonable approximation for this process \cite{Djouadi:2005gi}, and therefore we omit the bottom loop contribution in Eq. \eqref{kggh2}. While this analysis is conducted at LO, it is important to note that higher-order QCD corrections can induce significant effects \cite{PhysRevLett.79.2184, Chetyrkin_1998,Manohar_2006,deFlorian:2009hc}. For instance, the NLO QCD corrections, along with their precise quark mass dependence are well-known \cite{Inami:1983aa,Djouadi:1991tka,SPIRA199517}. They enhance the $H\rightarrow gg$ decay rate by approximately 70\%,  making them non-negligible. The QCD corrections beyond the NLO have been calculated in the heavy top quark limit \cite{Chetyrkin:1997iv,Baikov:2006ch,Herzog:2017dtz}, and they are less than 20\% of the NLO QCD-corrected $H\rightarrow gg$ decay \cite{DiMicco:2019ngk}. Furthermore, the NLO electro-weak corrections have been obtained for top and light quark loops \cite{Chetyrkin:1996wr,Djouadi:1994ge,Chetyrkin:1996ke}, and including virtual mass dependences in top, $W$ and $Z$ loop contributions \cite{Aglietti:2004nj,Aglietti:2006yd,Degrassi:2004mx}. 

With regards to the $H\rightarrow Z\gamma$ decay, the new quarks also contribute to the $h_1^{Z\gamma}$ form factor and modifies the signal strength $\mu^{Z\gamma}$ as follows \begin{equation}
\label{muZgamma2}
\mu^{Z\gamma}=1+(9.89\text{ GeV}^{-2})\delta \big|h_1^{Z\gamma}\big|^2+(1.94\text{ GeV}^{-2})\big|h_3^{Z\gamma}\big|^2,
\end{equation}
where $\delta \big|h_1^{Z\gamma}\big|^2$ is defined as in the $H\rightarrow \gamma\gamma$ decay. 

 
 \subsubsection{Numerical results}

To determine the allowed values for $\kappa$ and the masses of the new quarks, we investigate parameter regions where the presence of the heavy quarks does not lead to significant deviations in the observed signal strength $\mu^{\gamma\gamma}$, as well as the coupling modifier $\kappa_g$. It is found that for $\kappa\leqslant 0.05$, the contributions of the $T$ and $D$ quarks will be suppressed. For these values, the experimental results agree with $\mu^{\gamma\gamma}$ and $\kappa_{g}$ in Eq. \eqref{HgammaSignal2} and \eqref{kggh2} at 95\% CL. For the $H\rightarrow Z\gamma$ decay, we observed that the $\delta h_1^{Z\gamma}$ is approximately $\kappa\times 10^{-3}$, which is negligible compared to the LO contribution of the SM for any $\kappa< 1$. 
\begin{table}[H] \begin{center}
 \begin{tabular}{ccc|cc} Scenario& ${\rm Im }\left(K_A^{\overline{T}t}{K_{S}^{\overline{T}t}}^\ast\right)$  & ${\rm Im }\left(K_V^{\overline{T}t}{K_{P}^{\overline{T}t}}^\ast\right)$ & ${\rm Im }\left(K_A^{\overline{D}b}{K_{S}^{\overline{D}b}}^\ast\right)$& ${\rm Im }\left(K_V^{\overline{D}b}{K_{P}^{\overline{D}b}}^\ast\right)$ \\\hline $I$ &  1.8& 1.6  & 0.3   & 0.1   \\\hline $II$ & 0.7   & 0.9   &0.2  & 0.4   \\\hline $III$ & -1.8   & 1.6 & -0.3   & 0.1 \\\hline $IV$ & 0.7   &- 0.9   &0.2  &- 0.4  \end{tabular}\caption{New physics scenarios for the different combinations of imaginary parts, which contribute to the $CP$-violating form factor $h_3^{Z\gamma}$.}\label{scenarios}\end{center}
 \end{table}
  We are particularly interested in high values of the $CP$-violating form factor $h_3^{Z\gamma}$, as they may account for the observed excess in the $\mu^{Z\gamma}$ signal strength. These new contributions have to align closely with the central value in Eq. \eqref{signalLHC}, and for this reason, we will consider results consistent with the reported $\mu^{Z\gamma}$ by the LHC at the 68\% CL. Moreover, the masses of the new heavy quarks do not exceed 600 GeV, ensuring that the couplings of the new quarks will be within the perturbative limit and that the model will remain unitary \cite{CHANOWITZ1978285}. To find allowed masses of the $T$ and $D$ quarks, we present in Table \ref{scenarios} the scenarios considered for the different imaginary parts involved in the form factor $h_3^{Z\gamma}$.   In scenario $I$, the $\overline{D}b$ couplings are of order $10^{-1}$, while those associated with the up quarks are an order of magnitude larger. We also impose the condition ${\rm Im}\big(K^{\overline{Q}q}_A K^{\overline{Q}q}_S\big)>{\rm Im}\big(K^{\overline{Q}q}_V K^{\overline{Q}q}_P\big)$ for both up and down quark FCNC couplings. In contrast, scenario $II$ considers the opposite condition, where this inequality does not hold, indicating a different coupling hierarchy. In this scenario, the imaginary parts of the down and up quark couplings are of similar order of magnitude.
 
Both scenarios $I$ and $II$ restrict the analysis to positive values of the couplings, while the remaining cases incorporate negative imaginary parts to explore their impact on the behavior of $h_3^{Z\gamma}$. Throughout all scenarios, it is assumed that the magnitude of the up-quark couplings exceeds those of the down-quark couplings.
  
\begin{figure}[H]
\begin{center}
\includegraphics[width=13.5cm]{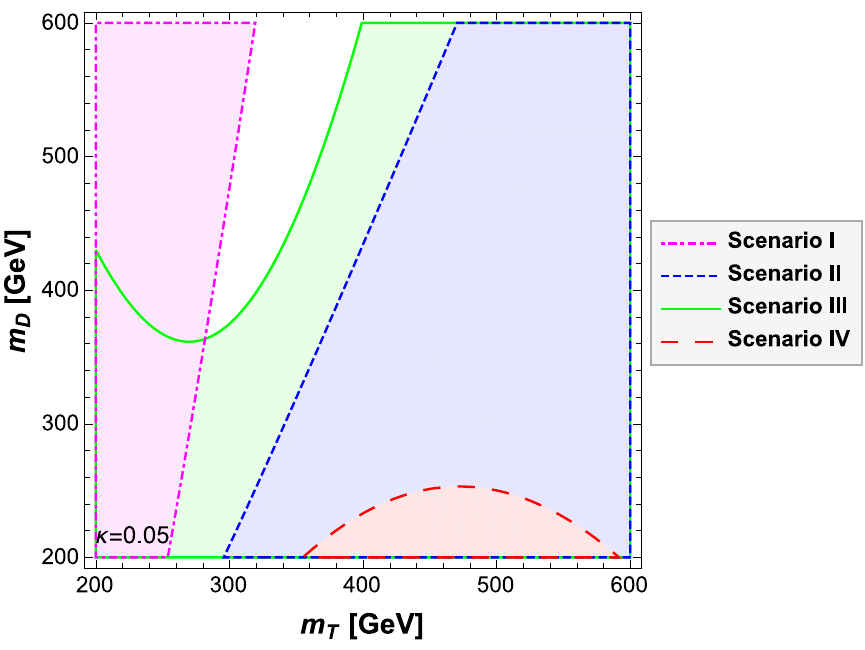}
\caption{Allowed regions for the masses of the new quarks that agree with $\mu^{Z\gamma}$ at 68\% of CL reported by the LHC. The contributions arising from the new quarks to the $H\rightarrow\gamma\gamma$ decay and $gg\rightarrow H$ production agree with $\mu^{\gamma\gamma}$ and $\kappa^{gg}$ at 95\% of CL.}\label{nf4}
\end{center}
\end{figure}

   In Fig. \ref{nf4}, we show the allowed areas in the $m_T$-$m_D$ plane for the scenarios outlined in Table \ref{scenarios}, which are consistent with the observed excess reported by the LHC in the $H\rightarrow Z\gamma$ channel at 68 \% of CL \cite{ATLAS:2016neq}, with $\kappa$ set to 0.05. In scenarios $I$-$III$, the case $m_D>m_T$ is possible in certain regions. This behavior is particularly evident in scenario $I$, where the mass of the $D$ quark can reach the largest values, with the $m_T$ mass restricted to 320 GeV. In contrast, in scenario $IV$, only high values of $m_T$ are permissible, and the $m_D$ can attain values up to 250 GeV. Our analysis reveals that including the minus sign in the imaginary parts leads to a noteworthy change in the behavior of the allowed regions, as observed in scenarios $III$-$IV$.

\begin{figure}[H]
\begin{center}
{\includegraphics[width=9.4cm]{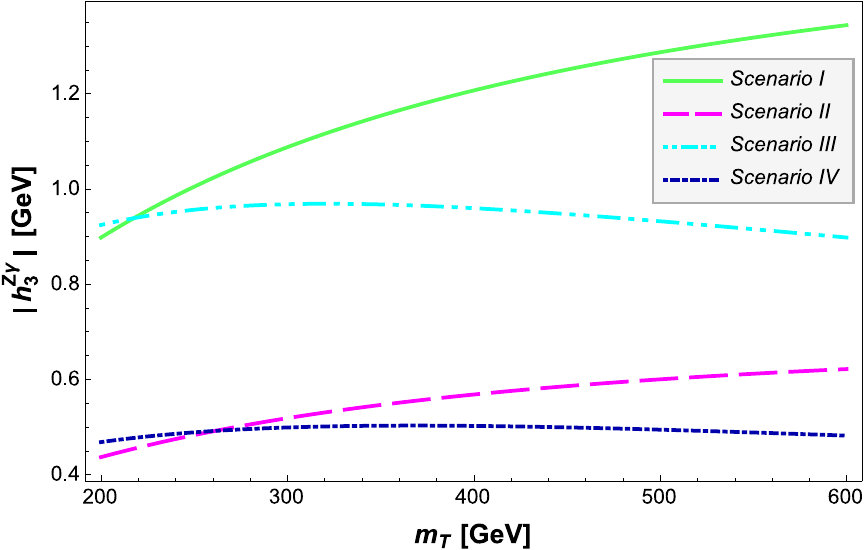}}
\caption{Contributions to $h_3^{Z\gamma}$ as function of $m_T$ for the four scenarios in Table \ref{scenarios}. We have used a mass splitting of 100 GeV.}\label{h3nc}
\end{center}
\end{figure}

\begin{figure}[H]
\begin{center}
{\includegraphics[width=9.4cm]{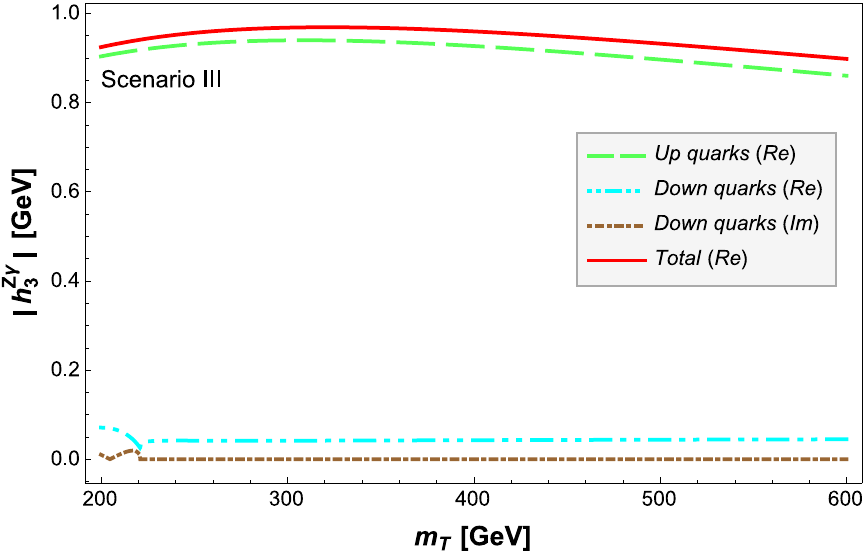}}
\caption{Real (Re) and imaginary (Im) contributions of the up and down quarks in scenario $III$ as a function of $m_T$. The total contribution is also shown. We have used a mass splitting of 100 GeV.}\label{h3nc2}
\end{center}
\end{figure}

To assess the new contribution to $h_3^{Z\gamma}$, we set the following mass difference between the $T$ and $D$ quarks 
\begin{equation}
m_T-m_D=100 \text{GeV}.
\end{equation}
  In Fig. \ref{h3nc}, we present the total contribution to $h_3^{Z\gamma}$ as a function of $m_T$ for the four scenarios in Table \ref{scenarios}. It is observed that the most significant values are obtained in scenarios $I$ and $III$, where high values of the up-quark couplings were considered. In scenario $I$, the magnitude $h_3^{Z\gamma}$ can exceed the unity, whereas, in scenario $III$, it approaches one. For the remaining scenarios, with relatively small and comparable up and down quark couplings, the magnitude of the $CP$-violating form factor can fall in the range $0.4\lesssim h_3^{Z\gamma}\lesssim 0.6$. It is important to note that within the permissible regions for the masses of the $T$ and $D$ quarks, our results are consistent with the upper bounds in Eq. \eqref{limh3}. In Fig \ref{h3nc2}, we show the individual contributions of the up and down quarks in scenario $III$, considering the allowable mass values from Fig. \ref{nf4}. It is worth noting that the up quarks only contribute to the real part of $h_3^{Z\gamma}$ and represent almost the total contribution. On the other hand, the down quarks can induce an imaginary part to $h_3^{Z\gamma}$ in regions where the $H\rightarrow \overline{D}b$ decay is kinematically permitted. Nevertheless, these corrections are of order $10^{-2}$, which is relatively small compared to those arising from the up quarks. Furthermore, we note a destructive interference between the up and down contributions, resulting in the total magnitude of $h_3^{Z\gamma}$ being lower than that of the up quarks. Similar behavior is found in the other scenarios.
 
\section{Conclusions}\label{Concsec}

In this work, we have studied new contributions to the $H\rightarrow Z\gamma$ decay. The $HZ\gamma$ vertex function is described by two form factors: one that is $CP$-conserving  ($h_1^{Z\gamma}$) and one that violates $CP$ symmetry ($h_3^{Z\gamma}$). We have calculated the $h_1^{Z\gamma}$ form factor in terms of the Passarino-Veltman scalar functions and confirmed our results with those reported in the literature. The dominant contributions to the real part of $h_1^{Z\gamma}$ come from $W$ boson loops, and it is of order $10^{-2}$ GeV, while the imaginary part is induced by a bottom quark and lepton loop corrections, which is of order $10^{-4}$ GeV. 

As the recent LHC result of $\mu^{Z\gamma}=2.2\pm 0.07$ for the $H\rightarrow Z\gamma$ signal strength, we investigated the possibility that the deviation from the SM prediction arises solely from $CP$ violation. We found that the real and absorptive parts of $h_3^{Z\gamma}$ can be less than 1.15 GeV at 95\% CL. Furthermore, we calculated the contribution to $h_3^{Z\gamma}$ from  FCNC mediated by the $Z$ and $H$ bosons with complex couplings and only SM particles. Our result is also expressed in terms of the Passarino-Veltman scalar functions. To estimate the magnitude of $h_3^{Z\gamma}$, we used limits on the $H\overline{t}c$ couplings, which were obtained through current LHC data on $t\rightarrow Hc$ decays and are of order $10^{-1}$ GeV. While for the $Z\overline{t}c$ couplings, we considered existing bounds in the literature. In this generic model, $h_3^{Z\gamma}$ can be up order $10^{-5}$ GeV when considering the top and charm quarks in the loop.
 
We examined a hypothetical model involving a new up quark $T$ and down $D$ quarks and included a factor $\kappa$ in the diagonal Higgs couplings with the new quarks. We find that for $\kappa\leqslant0.05$, the new contributions to the $H\rightarrow\gamma\gamma$ and $H\rightarrow gg$ decays are significantly suppressed and do not deviate the signal strength $\mu^{\gamma\gamma}$ and coupling modifier $\kappa_{g}$ from their observed results at the LHC. In addition, we investigated different scenarios of the couplings involved in the $h_3^{Z\gamma}$ form factor and identified the permitted masses of the new family of quarks that align with LHC results for $\mu^{Z\gamma}$. We obtain contributions to $h_3^{Z\gamma}$  of order $10^{-1}$ GeV, which may contribute to the excess on the signal strength $\mu^{Z\gamma}$.

 \begin{acknowledgments}
This work  was supported by UNAM Posdoctoral Program (POSDOC). We also acknowledge support from  Sistema Nacional de Investigadores (Mexico). 
\end{acknowledgments}

\appendix
\section{Bounds on $H\overline{t}c$ couplings}\label{ABoundHtc}
\begin{figure}[H]
\begin{center}
\includegraphics[width=8cm]{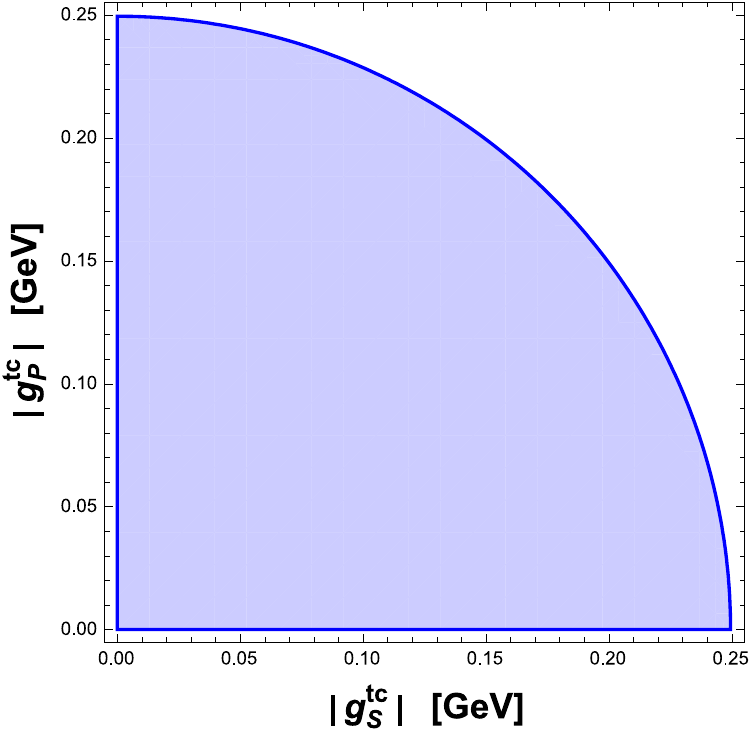}
\caption{Bounds on $H\overline{t}c$ couplings using the current constraint on $\mathcal{B}(t\rightarrow Hc)$ at the 95\% C.L. from the ATLAS collaboration \cite{ATLAS:2023ujo}.}\label{Htc}
\end{center}
\end{figure}

For FCNC couplings of the Higgs boson, bounds can be obtained using the reported limit from the ATLAS collaboration at the 95 \% CL, where $\mathcal{B}(t\rightarrow Hc)\leqslant5.8\times 10^{-4}$  \cite{ATLAS:2023ujo}. Moreover, from Lagrangian \eqref{lagFCNC}, the partial decay width for the $t\rightarrow Hc$ process can be computed as follows:
 \begin{equation}
\label{WidthHtc}
\Gamma(t\rightarrow Hc)=\frac{3 g^2 (m_H^2-m_t^2)^2}{64 \pi m_W^2 m_t^2}\Big\{ |g_S^{tc}|^2 +|g_P^{tc}|^2\Big\}.
\end{equation}
 Finally, constraints on the $g_S^{tc}$ and $g_S^{tc}$ couplings can be derived by combining both results. They are shown in Fig. \ref{Htc} and can be summarized as
 \begin{equation}
|g_{S,P}^{tc}|\lesssim 0.25 \ \text{GeV}.
\end{equation}

\section{Electro-weak contributions to the $H\rightarrow \gamma\gamma$ decay}\label{appendix1}

Following the definition of the $HZ\gamma$ vertex function in Sec. \ref{intro}, we write
\begin{equation}
\Gamma_{H\gamma\gamma}^{\mu\nu}=h_1^{\gamma\gamma} g^{\mu\nu}+\frac{1}{m_H^2} h^{\gamma\gamma}_2 p_1^\nu p_2^\mu,
\end{equation}
where we only consider $CP$-conserving form factors. Because of gauge invariance, they are related as follows 
\begin{equation}
h_2^{\gamma\gamma}=-2\ h_1^{\gamma\gamma}.
\end{equation}
The one-loop SM contributions to $h_1^{\gamma\gamma}$ have been studied in Refs. \cite{ELLIS1976292,Shifman:1979eb,Okun1982LeptonsAQ,GAVELA1981257}, and are similar to the presented in Fig. \ref{SMContributions}. The $W$ bosons contribution in terms of the Passarino-Veltman scalar functions is
\begin{align}
\label{h1Wgamma}
h_1^{\gamma\gamma}(W)=\frac{ e^2}{16 \pi ^2 m_W}  \Big\{-6 m_W^2
   \left(m_H^2-2 m_W^2\right)
   \text{C}_0\left(0,0,m_H^2,m_W^2,m
   _W^2,m_W^2\right)+m_H^2+6
   m_W^2\Big\},
\end{align}
whereas the fermion contribution is given by
\begin{align}
\label{h1Fgamma}
h_1^{\gamma\gamma}(F)=-\frac{ e^2 \mathcal{Q}_f^2 m_f^2\ N_c }{8 \pi
   ^2 m_W}
   \Big\{2-\left(m_H^2-4
   m_f^2\right)
   \text{C}_0\left(0,0,m_H^2,m_f^2,m
   _f^2,m_f^2\right)\Big\}.
\end{align}
Our results agree with those reported in Refs. \cite{Bergstrom:1985hp,Fortes:2014dia}. In the SM, the numerical values are
\begin{equation}
\label{totgamma}
h_1^{\gamma\gamma}(W)=0.48\ \text{GeV,}\quad h_1^{\gamma\gamma}(F)=-(0.103+i0.0004)\ \text{GeV},
\end{equation}
where for the fermion contribution we considered the top and bottom quarks and the $e^-$, $\mu^-$, and $\tau^-$ leptons. The two-loop QCD corrections have been reported in Refs. \cite{PhysRevD.42.3760,DJOUADI1991187,PhysRevD.47.1264,Melnikov_1993,DJOUADI1993255,S0217732394001003,Fleischer:1994aa,SPIRA199517,FLEISCHER2004294,Harlander2005,Anastasiou_2007,Aglietti_2007}. The NLO electro-weak corrections have been also calculated numerically in Refs \cite{ACTIS2009182,DJOUADI199817,DEGRASSI2005183,PASSARINO2007298}. Moreover, the beyond NLO QCD corrections were estimated in the heavy top quark limit \cite{steinhauser1996correc,Sturm_2014,MAIERHOFER2013131}. 

\section{$gg\rightarrow H$}\label{appendix2}
The main channel of Higgs production is through gluon-gluon fusion. It is proportional to the $H\rightarrow gg$ width, which is given as \cite{Gunion:1989we,Djouadi:2005gi,Angelescu:2015uiz}
\begin{equation}
\sigma(gg\rightarrow H)\propto \Gamma(H\rightarrow gg)=\frac{G_\mu\alpha_s^2M_H^3}{64\sqrt{2}\pi^3}\Bigg|\frac{3}{4}\sum_fA_f(\tau_f)\Bigg|^2,
\end{equation}
where $A_f$ is the form factor for spin-1/2 particles, and is given as follows
\begin{equation}
A_f(\tau)=2\big[\tau+(\tau-1)f(\tau)\big]\tau^{-2},
\end{equation}
with $\tau_f=m_H^2/4m_f^2$ and the three-point integral $f(\tau)$ defined as
\begin{equation}
f(\tau)=\left\{\begin{array}{cc}\arcsin^2\sqrt{\tau} & \tau\leqslant1 \\ -\frac{1}{4}\Big[\ln\frac{1+\sqrt{1-\tau^{-1}}}{1-\sqrt{1-\tau^{-1}}}-i\pi\Big]^2 & \tau>1\end{array}\right.
\end{equation}

\bibliography{BiblioH}
\end{document}